\documentclass[a4paper,pdflatex]{article}
\usepackage{graphicx,amsmath,amsfonts}
\RequirePackage[english]{babel}
\usepackage[T2A]{fontenc}
\usepackage[cp1251]{inputenc}
\usepackage{enumerate}
\title{\Large \bf Corrections to the higher moments of the heavy ion energy-loss distribution beyond the Born approximation. \\II. $\bf{\beta}$ dependence of Mott's corrections}
\author{
O. Voskresenskaya}
\date{}

\begin{document}
\maketitle

\begin{center}
Joint Institute for Nuclear Research, Dubna, Moscow Region, 141980 Russia
\end{center}

\begin{abstract}
The Mott corrections to the higher moments of
the heavy ion energy-loss distribution are calculated
in a wide range of relative particle velocity on the basis of the Mott exact cross section. It is shown that the relative Mott corrections to the first-order Born central moments and normalized central moments reach a large value over the range under consideration.

\end{abstract}


\section{Introduction}

The average energy loss of a relativistic particle per unit path length
is described by the relativistic version of the Bethe formula \cite{Bethe1, Bethe3}

\begin{equation}
\label{Bethe}
-\frac{d \bar E}{dx}=2\zeta\left[\ln\left(\frac{E_m}{I}\right)-\beta^2\right]\nonumber
\end{equation}
with
\begin{equation}
\zeta=K\left(\frac{Z}{\beta}\right)^2\frac{Z^\prime}{A},\quad K=2\pi N_A\frac{e^4}{mc^2},\quad
E_m\approx \frac{2m c^2\beta^2}{1-\beta^2}\,.
\end{equation}
Here, $x$ is the distance traveled by a particle,
$E_m$ denotes the maximum transferrable energy to an electron in a collision with the particle
of velocity $\beta c$, $m$ is the electron mass,
$Z^\prime$ and $A$  are the atomic number and the weight of an absorber, respectively, $N_A$ is the Avogadro number, and $I$
is its mean excitation potential.

The importance of the Mott higher-order-correction term $\Phi_M/2$ \cite{Mott} to the formula (1),
\begin{eqnarray}
\label{BetheM}
-\frac{d\bar E}{dx}=2\zeta\left[\ln\left(\frac{E_m}{I}\right)-\beta^2 +\frac{\Phi_M}{2}\right]\,,\nonumber
\end{eqnarray}
\begin{equation}
\Phi_M = \frac{1}{\zeta}\Delta_M\!\!\left(\frac{d\bar E}{dx}\right),\quad
\Delta_M\!\!\left(\frac{d\bar E}{dx}\right)=N\!\!\int\!\!\left(\frac{d\sigma_M}{d\varepsilon}-
\frac{d\sigma_B}{d\varepsilon}\right)\!\varepsilon d\varepsilon,
\label{Bethe3}
\end{equation}
was noted in various works (see, e.q., \cite{GSI1}). In the above formula
$\Delta_M$ denotes the Mott correction (MC),
$N=N_A Z^\prime/A$ is the number of target electrons per unit volume,  and $\sigma_B$
represents the first-order Born approximation to the Mott exact cross section $\sigma_M$.

The expression for the MCs in (\ref{Bethe3}) is extremely inconvenient for practical application. In this regard, obtaining convenient and accurate representations for the  MCs become significant.
Ref. \cite{VSTT} gives an exact expression for the Mott correction  in the form of a
rather fast converging series of the quantities bilinear in the Mott partial amplitudes, which can be  quite simply calculated.

The aim of the presented work is to estimate
the relative Mott corrections  to the first-order Born central moments and normalized central moments of the energy-loss distribution for some heavy ions in a wide range of relative particle velocity based on results of \cite{VSTT}.
The presented communication is organized as follows. Section 2 considers
the analytical expressions for the higher moments of the heavy ion energy-loss distribution
in the form of a quite rapidly converging series on the basis of \cite{VSTT}.
Section 3 gives the results of numerical calculations for the relative  Mott corrections to  the n-th central moments of the distribution. Section 4 presents numerical results for the relative  Mott corrections to  the normalized distribution k-th central moments.
In Section 5 we briefly sum up our results and outline some prospects.
This paper is a continuation of our communication \cite{preprint1}, where the $Z$ dependence of the Mott corrections to the energy-loss distribution moments was examined.

\section{Analytical expressions for the higher moments of energy-loss distribution}

\smallskip

Let us consider the central moments of the relativistic ion energy-loss distribution  in the Mott approximation:
\begin{equation}
\label{mun}
\mu_{n,M}=2\pi N\Delta x\int\limits_{0}^{\pi}[\Delta\varepsilon(\vartheta)]^{n}
\frac{d\sigma_M}{d\Omega}\sin\vartheta d\vartheta,\quad \Delta\varepsilon(\vartheta)=\frac{2p^2}{m}(1-\cos\vartheta)\,,
\end{equation}
where $\Delta x$ is the thickness of the target material traversed by the ion, $\Delta\varepsilon(\vartheta)$ is the energy transfer to a target electron in a collision
leading to a center-of-mass scattering angle
$\vartheta$, and
the Mott exact cross section
can be expressed as follows
\begin{equation}\frac{d\sigma_M}{d\Omega}=\frac{\hbar^2}{4p^2\sin^2(\vartheta/2)}
\Bigl[\xi^{2}\vert F(\vartheta)\vert^{2}+\vert G(\vartheta)\vert^2\Bigl],
\end{equation}
\begin{equation}
F(\vartheta)=\sum_{l=0}^{\infty}F_lP_l(x),~x=\cos \vartheta,\quad
G(\vartheta)=-cos(\vartheta/2)dF(\vartheta)/d\vartheta\, ,
\end{equation}
\begin{equation}
F_l=lC_l-(l+1)C_{l+1},\quad
C_l=\frac{\Gamma(\rho_l-i\nu)}{\Gamma(\rho_l+1+i\nu)}e^{i\pi(l-\rho_l)},\end{equation}
\begin{equation}
\label{sigma}
p=mc\frac{\beta}{\sqrt{1-\beta^2}},\quad\xi=\nu\sqrt{1-\beta^2},\quad
\rho_l=\sqrt{l^2-(Z\alpha)^2},\quad \nu=\frac{Z\alpha}{\beta}\,.
\end{equation}
Here, $P_l$  is the Legendre polynomial of order $l$
and $\Gamma(\mu)$ designates the Euler gamma function.

With the use of the  expression
\begin{eqnarray}
(2l+1)xP_{l}^{m}(\cos \vartheta)&=&(l+1-\vert m\vert)P_{l+1}^{m}(\cos \vartheta)-(l+\vert m\vert)P_{l-1}^{m}(\cos \vartheta)
\end{eqnarray}
and the orthogonality relation
\begin{equation}
\int\limits_{-1}^{1}P_{l_1}^{m}(\cos \vartheta)P_{l_2}^{n}(\cos \vartheta)\sin \vartheta d\vartheta=\frac{2}{2l_1+1}\,
\frac{(l_1+\vert m\vert) !}{(l_1-\vert m\vert)!}\delta_{l_1l_2}\,,
\end{equation}
we can obtain from (3)--(7) the following representations for the quantities
$\mu_{n,M}\; (n=\overline{2,4})$ in the form of rapidly converging series:
\begin{eqnarray}
\label{mu2}
\mu_{2,M}=a\sum\limits_{l=0}^{\infty}(l+1)\left[\xi^2
\left\vert\frac{F_l}{2l+1}-\frac{F_{l-1}}{2l+3}\right\vert^2
+\left\vert\frac{lF_l}{2l+1}-\frac{(l+2)F_{l+1}}{2l+3}\right\vert^2\right]\,,
\end{eqnarray}
\begin{eqnarray}
\label{mu3}
\mu_{3,M}=\frac{ap^2}{m}\sum\limits_{l=0}^{\infty}\frac{1}{(2l+1)}
\Bigl[\xi^2\big\vert\tilde F_l\big\vert^2+l(l+1)\big\vert\tilde G_l\big\vert^2\Bigl]\,,
\end{eqnarray}
\begin{eqnarray}
\label{mu4}
\mu_{4,M}=a\left(\frac{p^2}{m_e}\right)^2\sum\limits_{l=0}^{\infty}(l+1)
\left[\xi^2\left\vert\frac{\tilde F_l}{2l+1}-\frac{\tilde F_{l+1}}{2l+3}\right\vert^2
+\left\vert\frac{l\tilde G_l}{2l+1}-\frac{(l+2)\tilde G_{l+1}}{2l+3}\right\vert^2\right]
\end{eqnarray}
with
\begin{eqnarray}
\tilde F_l=F_l-\frac{l}{2l-1}F_{l-1}-\frac{l+1}{2l+3}F_{l+1},\quad \tilde G_l=F_l-\frac{l-1}{2l-1}F_{l-1}-\frac{l+2}{2l+3}F_{l+1}\,,
\end{eqnarray}
\begin{eqnarray}
\label{a}
a=\frac{2\pi N\Delta x(\hbar c\beta)^2}{1-\beta^2}\,.
\end{eqnarray}\\
The terms of these series decrease
as $l^{-2n+1}~(n=\overline{2,4})$ when $l\to \infty$, and that the series converges absolutely.

\section{Numerical results for the MCs to the central moments of the heavy ion energy-loss distribution}

\bigskip

The dominant contribution to the higher moments of the particle energy-loss distribution that determine its shape \cite{IDDRS} is made by close collisions. Therefore,
we can expect significant deviations from the results of the Born approximation in calculations
of these quantities.

\bigskip

Let us introduce the quantities that characterize the relative Mott corrections in
computing the distribution  central moments $\mu_{n}$ ($n=\overline{2,4}$):
\begin{equation}
\delta_n\equiv\delta_n(\mu)=\frac{\mu_{n,M}-\mu_{n,B}}{\mu_{n,B}},\quad n=\overline{2,4},
\end{equation}

\medskip

\noindent where  $\mu_{n,M}$ values
($n=\overline{2,4}$) are determined by
(\ref{mu2})--(\ref{a}), and $\mu_{n,B}$ ($n=\overline{2,4}$) are given by \\

\begin{equation}
\mu_{n,B}=\pi N\Delta x \left(\frac{\nu \hbar}{m}\right)^2
\left(\frac{2p^2}{m}\right)^n\left(\frac{1}{n-1}-\frac{\beta^2}{n}\right)\,.
\end{equation}\\

The calculation results for the above corrections $\delta_n(Z,\beta)$
are given in Table 1 and are also illustrated in Fig. 1; they
give the dependence of $\delta_n$  on the particle relative velocity $\beta=v/c$.\\\\

\begin{figure}[h!]

\begin{center}

\includegraphics[width=0.7
\linewidth]{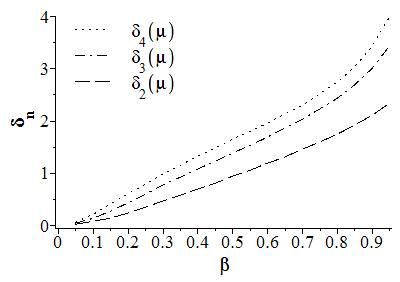}

\caption{Relative corrections  $\delta_n$ to the Born central moments
$\mu_{n,B}$ ($n=\overline{2,4}$) of the heavy ion energy-loss distribution depending on the relative ion velocity $\beta$ for $Z = 92$.
} \label{Fig2}

\end{center}

\end{figure}

\medskip

It can be seen from Fig. 1 and Table 1 that the relative Mott
corrections $\delta_n(\mu)$ can reach a very large value.
So the Mott energy-loss straggling
$\mu_{2,M}$
is larger than that in the Born approximation $\mu_{2,B}$ by a
factor of 3 for $Z = 92$ ($R=\mu_{2,M}/\mu_{2,B}=\delta_2(\mu)+1\sim 3$),
which is consistent with the results of the ELS measurements for $Z = 92$ \cite{GSI1}.
The magnitudes of $\mu_{n,M}$ ($n=\overline{3,4}$)  in the Mott approximation
are larger than those in the Born approximation $\mu_{n,B}$ ($n=\overline{3,4}$) by a
factors of 4 and 5, correspondingly, for $Z = 92$.

\newpage

\begin{center}
\noindent  {\bf Table 1.} $\beta$ dependence
of the relative corrections  $\delta_n(\mu)$
to the first-order Born central moments $\mu_{n,B}$
($n=\overline{2,4}$) for $Z=92$.

\medskip
\begin{tabular}{|c|c|c|c|}
\hline
$\beta$& $\delta_2(\mu)$& $\delta_3(\mu)$& $\delta_4(\mu)$\\
\hline
0.0500& 0.0155& 0.0362& 0.0570\\
0.1500& 0.1383& 0.2675& 0.3879\\
0.2500& 0.3336& 0.5877& 0.7841\\
0.3500& 0.5617& 0.9129& 1.1478\\
0.4500& 0.8016& 1.2242& 1.4766\\
0.5500& 1.0481& 1.5290& 1.7893\\
0.6500& 1.3055& 1.8467& 2.1137\\
0.7500& 1.5865& 2.2105& 2.4950\\
0.8500& 1.9140& 2.6838& 3.0261\\
0.9500& 2.3303& 3.4152& 3.9685\\
\hline
\end{tabular}

\end{center}

\bigskip

Thus, we can conclude that the Born approximation can not be used to adequate estimation of the
higher moments of the heavy ion energy-loss distribution. Their estimation requires going beyond this approximation using, in particular, Mott's corrections.

\section{Numerical results for the Mott corrections to  the normalized k-th central moments}

The relative Mott corrections to the normalized 3rd and 4th central moments $\rho_{k,B}$ ($k=\overline{3,4}$) of the relativistic ion energy-loss distribution may be defined as follows:

\begin{equation}
\delta_k\equiv\delta_k(\rho)=\frac{\rho_{k,M}-\rho_{k,B}}{\rho_{k,B}},\quad k=\overline{3,4}
\end{equation}
where
\begin{eqnarray}
\rho_{3,M}=\frac{\mu_{3,M}}{\big(\mu_{2,M})^{3/2}}\,,\qquad
\rho_{4,M}=\frac{\mu_{4,M}}{\big(\mu_{2,M})^{4/2}}\,.
\end{eqnarray}

\noindent Let us evaluate these corrections some highly charged particles (e.q., for  $_{92}$U ions). Figure 2 demonstrates the numerical results of the  $\delta_k(\rho)$ calculations and shows the dependence of these corrections on the relative particle velocity $\beta$
 for $Z = 92$. Table 2 listed their values over the range $0.35\leq\beta \leq 0.95$. \\

\begin{figure}[h!]
\begin{center}
\includegraphics[width=0.65\linewidth]{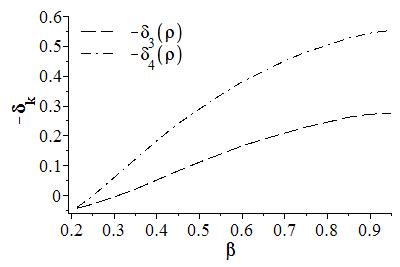}
\caption{\small Relative corrections  $\delta_k$ to the normalized Born central moments
$\rho_{k,B}$ ($k=\overline{3,4}$) of the energy-loss distribution of heavy
charged particles depending on the relative particle velocity $\beta$
 for $Z = 92$.
} \label{Fig4}
\end{center}
\end{figure}

\begin{center}
{\bf Table 2.}  $\beta$ dependence of the relative corrections $\delta_k(\rho)$
to the normalized first-order Born central moments $\rho_k$
($k=\overline{3,4}$) for $Z=92$.

\medskip

\begin{tabular}{|c|r|r|}
\hline
$~~\beta~~$&  $-\delta_3(\rho)~~$& $-\delta_4(\rho)~~$\\
\hline
~~0.3500~~&  $ 0.0198~~$& $ 0.1193~~$\\
~~0.4500~~&  $ 0.0802~~$& $ 0.2370~~$\\
~~0.5500~~&  $ 0.1371~~$& $ 0.3350~~$\\
~~0.6500~~&  $ 0.1868~~$& $ 0.4142~~$\\
~~0.7500~~&  $ 0.2282~~$& $ 0.4776~~$\\
~~0.8500~~&  $ 0.2594~~$& $ 0.5259~~$\\
~~0.9500~~&  $ 0.2735~~$& $ 0.5520~~$\\
\hline
\end{tabular}
\end{center}

The modules of obtained relative Mott corrections to  the normalized k-th central moments vary
between 2--12$\%$ percent for $\beta=0.35$ and 27--55$\%$ for $\beta=0.95$ over the range considered. The results obtained mean that the heavy ion energy-loss distributions, which  were
computed with the use of Mott's corrections, are less asymmetric and more close to Gaussian
that those in the Born approximation.

\section{Summary and outlook}

Based on the representation of the Mott corrections $\Delta_\textrm{M} $ to the Bethe-Bloch formula in the form of rapidly convergent series of quantities bilinear in the Mott partial amplitudes,
we managed to reduce the calculation of the most important central moments $\mu_\textrm {n,M}$ ($n = \overline{2,4} $) of the average energy-loss distribution of heavy ions in the Mott approximation to the summation of rapidly convergent (as $ l^{-2n + 1} $ at $ l \to \infty $, $ l\in\mathbb{N}_0$) infinite series. As a result, we were able to show
that whereas the relative corrections $\delta_k$
($k=\overline{3,4}$) to the normalized first-order Born central moments $\rho_k$
($k=\overline{3,4}$) can reach several tens of percent over the $\beta$ range considered, these corrections achieve several hundred percent for the  $\mu_{n,M}$ ($n=\overline{2,4}$) over the range  $0.05\leq\beta \leq 0.95$. In particular, this explain results of experiment \cite{GSI1}. It is of interest to find the Coulomb corrections (CCs) to the higher moments of the energy-loss distribution and to estimate the total corrections to these quantities.
Comparison with the results of other experiments is planned later.

\section*{Acknowledgments}

This work was supported by a grant from the Russian Foundation for Basic Research
(project nos. 17-01-00661-а).


\end{document}